\documentclass[useAMS,usenatbib,usegraphicx]{mn2e}

\title[The planetary nebula IRAS\,18197$-$1118]{The planetary nebula nature
  and properties of IRAS\,18197--1118\thanks{Based on 
observations collected at the German-Spanish Astronomical Center, Calar Alto,
jointly operated by the Max-Planck-Institut f\"ur 
Astronomie (Heidelberg) and the Instituto de Astrofísica de Andalucía (CSIC).}\thanks{Based upon observations acquired at the 
Observatorio Astron\'omico Nacional in the Sierra San Pedro M\'artir (OAN-SPM), Baja California, Mexico.}}
\author[L.F. Miranda et al.]{L.\,F. Miranda$^{1,2}$\thanks{E-mail:
lfm@iaa.es (LFM); l.rodriguez@crya.unam.mx (LFR); claudio@on.br (CBP),
vazquez@astrosen.unam.mx (RV)}\thanks{LFM is Personal Desplazado of CSIC on
temporal leave at the University of Vigo}, 
  L.\,F. Rodr\'{\i}guez$^{3}$\footnotemark[3],
  C.\,B. Pereira$^{4}$\footnotemark[3], R. V\'azquez$^5$\footnotemark[3]\\
$^{1}$ Consejo Superior de Investigaciones Cient\'{\i}ficas, C/ Serrano 117,
E-28006 Madrid, Spain \\
$^{2}$ Departamento de F\'{\i}sica Aplicada, Facultade de Ciencias, Campus
Lagoas-Marcosende s/n,  Universidade de Vigo, \\ E-36310 Vigo, Spain (present address) \\
$^{3}$ Centro de Radioastronom\'{\i}a y Astrof\'{\i}sica, Universidad Nacional
Aut\'onoma de M\'exico, Apartado Postal 3-72, Morelia, 
Michoac\'an 58089, Mexico \\
$^{4}$ Observat\'orio Nacional, Rua Jos\'e Cristino, 77. CEP 20921-400, S\~ao
Crist\'ov\~ao, Rio de Janeiro-RJ, Brazil  \\
$^{5}$ Instituto de Astronom\'{\i}a, Universidad Nacional Aut\'onoma de M\'exico, Apdo. Postal 877, 
22800 Ensenada, BC, Mexico
}

\begin{document}

\date{Accepted . Received ; in original form }

\pagerange{\pageref{firstpage}--\pageref{lastpage}} \pubyear{}

\maketitle

\label{firstpage}

\begin{abstract} 
IRAS\,18197$-$1118 is a stellar-like object that has been
    classified as a planetary nebula from its radio continuum emission and high
    [S\,{\sc iii}]$\lambda$9532 to Paschen\,9 line intensity ratio, as derived
    from direct images. We present intermediate- and high-resolution, optical
    spectroscopy, VLA 8.46\,GHz radio continuum data, and 
    narrow-band optical images of IRAS\,18197$-$1118 aimed at confirming its
    planetary nebula nature, and analyzing its properties. The optical spectrum shows that
    IRAS\,18197$-$1118 is a medium-excitation planetary nebula suffering a
    high extinction ($c_{\rm H\beta}$ $\simeq$ 3.37). The
    optical images do not resolve the object but the 8.46\,GHz image reveals
    an elliptical shell of $\simeq$ 2.7$\times$1.6
    arcsec$^2$ in size, a compact central nebular region, and possible bipolar
    jet-like features, indicating several ejection events. The existence of a compact central nebula 
makes IRAS\,18197--1118 singular because this kind of structure  
      is observed in a few PNe only. An expansion velocity $\simeq$ 20\,km\,s$^{-1}$ and a systemic 
velocity (LSR) $\simeq$ +95\,km\,s$^{-1}$ are obtained for the object. An electron density of $\simeq$
    3.4$\times$10$^4$\,cm$^{-3}$ and an ionized mass of $\simeq$
    2.1$\times$10$^{-2}$\,M$_{\sun}$ are deduced from the 8.46\,GHz radio
    continuum data for an estimated statistical distance of 6\,kpc. Helium
    abundance is high but nitrogen is not enriched, which is not consistently reproduced 
    by evolutionary models, suggesting different abundances in the elliptical shell and central
    region. The properties of IRAS\,18197$-$1118 indicate a relatively young planetary nebula, 
    favor a distance of $\ga$ 6\,kpc, and strongly suggest that it is an inner-disc 
    planetary nebula.   
\end{abstract}

\begin{keywords}
planetary nebula: individual (IRAS\,18197$-$1118) -- interstellar medium: jets and 
outflows  -- interstellar medium: abundances
\end{keywords}

\section{Introduction}
Planetary nebulae (PNe) are the evolutionary phase of low- and
intermediate-mass stars ($M$ $\simeq$ 0.8--10\,M$_{\odot}$), between the
asymptotic giant branch and the white dwarf phase. They are key objects to
study phenomena taking place during the late evolution of low- and
intermediate-mass stars as, e.g., the mass 
ejection in the asymptotic giant branch, the formation of complex PNe, and the return of 
processed material to the interstellar medium, among others. In these studies, a precise
knowledge of the total number of PNe and their properties is
crucial. Remarkably, there exists a large discrepancy
between the known and predicted number of PNe 
in the Galaxy, being the predicted number larger or much larger than the known
one (see, e.g., Jacoby et al. 2010 and references
therein). In recent years, surveys in different wavelength ranges have provided the opportunity 
of identifying large amounts of new PN candidates (e.g., 
Helfand et al. 1992; Condon, Kaplan \& Terzian 1999; Parker et al. 2006; Miszalski et
al. 2008; Viironen et al. 2009; Jacoby et al. 2010; Ramos-Larios et al. 2012),
many of which have already been confirmed as true PNe by optical spectroscopy (e.g., Parker et
al. 2006). As for the optical surveys, it is expected that only very few PNe have been 
missed in the surveyed areas, although compact (stellar-like), very low
surface brightness, and/or highly extincted PNe may be difficult to recognize
(e.g., Miszalski et al. 2008). 
Identifying possible missing cases is interesting not only because they add to
the number of known PNe but also because they may help refine automatic
searches of PNe in large observational databases. 

IRAS\,18197$-$1118 [$\alpha$(2000.0) = $18^{\rm h}$ $22^{\rm m}$
$30\rlap.^{\rm s}0$, $\delta$(2000.0) = $-11^{\circ}$ 16$'$ 44$''$; $l$ =
$019\rlap.^{\circ}6095$, $b$ = $+01\rlap.^{\circ}1877$] is a stellar-like object that 
was identified as a PN candidate by Helfand et al. (1992) in an extension of
the Galactic plane survey at 20\,cm (Zoonematkermani et al. 1990). Subsequent 
radio continuum surveys have always detected the object. In particular, radio continuum emission from 
IRAS\,18197$-$1118 has been detected also at 1.4\,GHz with a flux density of 11.3$\pm$0.7\,mJy 
in the NRAO VLA Sky Survey (Condon et al. 1999), and at 
4.86 GHz with a flux density of 60.7$\pm$0.9\,mJy in the VLA Red MSX Source
(RMS) survey (Urquhart et al. 2009). 
Surprisingly, IRAS\,18197$-$1118 has never been identified in any optical/infrared survey 
neither has it been included in analysis of PNe and PN candidates, with the exception of the 
work by Kistiakowsky \& Helfand (1995, hereafter KH95). 
These authors did include IRAS\,18197$-$1118 in their list of highly extincted, presumably
distant PN candidates to be confirmed through the [S\,{\sc iii}]$\lambda$9532/Paschen\,9 line 
intensity ratio as derived from direct images in these two lines. The high
value of this ratio found in IRAS\,18197$-$1118, as compared with that in
H\,{\sc ii} regions, leads KH95 to propose a PN
nature for the object. Even though this result and the presence of radio continuum emission
favor that IRAS\,18197$-$1118 is a PN, an analysis of its optical spectrum
would be desirable to provide a firm confirmation of its nature, and to study
its physical conditions and chemical abundances. In addition, the published 
VLA observations at 1.4 and 4.86\,GHz (Condon et al. 1999; Urquhart et
al. 2009) do not have spatial resolution 
enough to resolve IRAS\,18197$-$1118 and, in consequence, its morphology is unknown. 

We have carried out an analysis of intermediate- and high-resolution, optical spectra,
VLA 8.46\,GHz radio continuum data, and narrow-band optical images of
IRAS\,18197--1118 with the aim of confirming its PN nature and studying its
properties. In this paper we present the results of this investigation. 

\section{Observations}

\subsection{Intermediate-resolution optical spectroscopy}

Intermediate-resolution, long-slit spectra of IRAS\,18197$-$1118 were obtained on Calar Alto 
Observatory (Almer\'{\i}a, Spain) with the Calar Alto Faint Object Spectrograph (CAFOS) 
at the 2.2\,m telescope on 2008 June 27. The detector was a SITe CCD with 2048$\times$2048 pixels. 
We used grisms B-100 and R-100 to cover the spectral ranges 3200--6200\,{\AA} and 5800--9600\,{\AA}, 
respectively, at a dispersion of 1.98\,{\AA}\,pixel$^{-1}$. The slit (2\,arcsec wide) was centered 
on the object and oriented at position angle (PA) 90$^{\circ}$. The exposure time was 2400\,s 
for each grism. The spectrophotometric standard star BD+28$^{\circ}$4211 was
also observed for flux calibration. Sky was photometric during the observations and seeing was 
$\simeq$ 1\,arcsec. The spectra were reduced following standard procedures within the {\sc midas} 
and {\sc iraf} packages. We note that the flux calibration fails at the
  very red edge of the CCD, coinciding with the [S\,{\sc iii}]$\lambda$9532
  emission line detected in the spectrum. As a result, the flux of this
  emission line is unrealistic and it will not be considered further in this paper.

\subsection{High-resolution optical spectroscopy}

High-resolution, long-slit spectra were obtained with the Manchester
Echelle Spectrometer (MES; Meaburn et al. 2003) at the 2.1\,m telescope of the 
OAN-SPM\footnote{The Observatorio Astron\'omico 
Nacional at the Sierra de San Pedro M\'artir, Baja California (OAN-SPM) 
is operated by the Instituto de Astronom\'{\i}a of the Universidad Nacional Aut\'onoma de M\'exico (IA-UNAM).} 
observatory during 2013 September 23.
An e2v CCD with 2048$\times$2048 pixels was used as
detector, in the 2$\times$2 binning mode (0.6 arcsec\,pix$^{-1}$ plate scale).
The slit length is 6.5\,arcmin and its width was set to 2\,arcsec. Spectra with the slit oriented 
at PA +72{\degr} and $-$18{\degr} were obtained using a $\Delta\lambda = 90$\,{\AA} bandwidth 
filter to isolate the 87$^{\rm th}$ order (0.1\,{\AA}\,pix$^{-1}$ dispersion),
covering the H$\alpha$ and [N\,{\sc ii}]$\lambda\lambda$6548,6583 emission
lines. In addition, a spectrum 
with the slit oriented at PA $-18${\degr} was obtained using a $\Delta\lambda = 50$\,{\AA} 
bandwidth filter to isolate the 114$^{\rm th}$ order (0.08\,{\AA}\,pix$^{-1}$ dispersion), 
covering the [O\,{\sc iii}]$\lambda$5007 emission line. Exposure time was 1800\,seg for 
each spectrum. Seeing was $\simeq$ 1.8\,arcsec during the observations. Data
were calibrated using standard techniques for long-slit spectroscopy 
in the {\sc iraf} package. The resulting spectral resolution (FWHM) is 
$\simeq12\,{\rm km\,s}^{-1}$ (accuracy $\pm$1\,km\,s$^{-1}$), as measured from the 
lines of the ThAr calibration lamp.

\subsection{Radio continuum observations}

The radio observations of IRAS~18197$-$1118 were obtained from the archive of
the Very Large Array (VLA)
of the NRAO\footnote{The National Radio Astronomy Observatory is operated by Associated Universities
Inc. under cooperative agreement with the National Science Foundation.}. These
unpublished archive observations were made 
at 8.46\,GHz in a snapshot of 8 minutes duration in 2005 January 19, under
project AC761. The array was then in the BnA configuration.
The data were edited and calibrated using the software package Astronomical
Image Processing System (AIPS) of NRAO.
The amplitude calibrator was 1331+305, with an adopted flux density of
5.21\,Jy and the phase calibrator was 1832-105,
with a bootstrapped flux density of 1.407$\pm$0.003\,Jy. The synthesized beam is 
0.51$\times$0.28\,arcsec$^2$ at PA $77^\circ$.

The integrated flux density of IRAS~18197$-$1118 at 8.46\,GHz is
64.9$\pm$0.6\,mJy. A comparison of the fluxes at
1.4, 4.86, and 8.46\,GHz indicates that the source is optically thin for
frequencies above 8.46\,GHz

\begin{figure*}
   \begin{center}
   \includegraphics[angle=-90,width=140mm,clip]{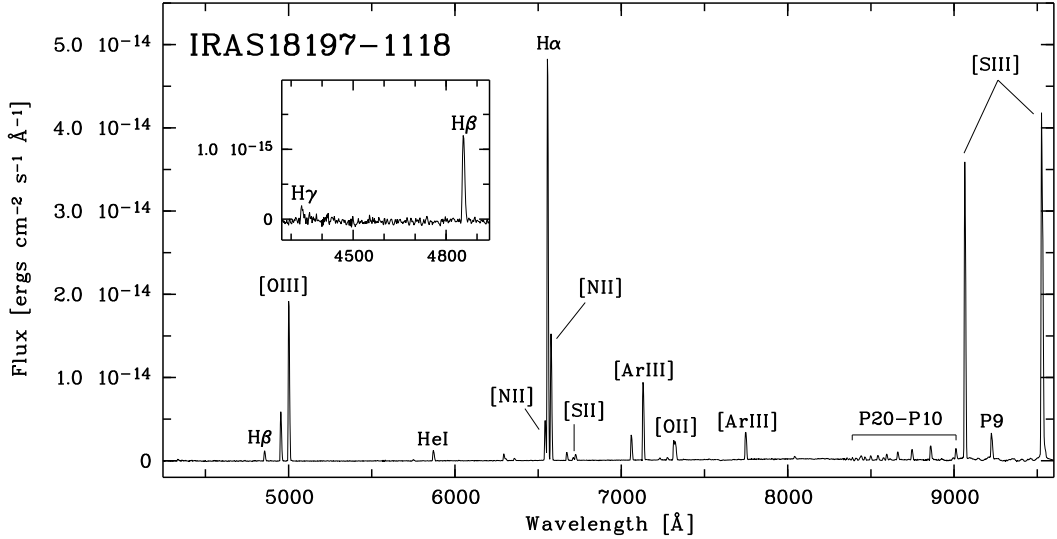}
   \caption{Blue and red CAFOS observed spectra of IRAS\,18197--1118 in the
     spectral range 4245--9480 {\AA}. Some emission lines are labelled. The
     inset shows the spectrum between the H$\gamma$ and H$\beta$ emission lines. }
    \end{center}
    \end{figure*}

\subsection{Optical images}

Narrow-band optical images of IRAS\,18197--1118 were obtained with the 1.5\,m telescope of 
the Observatorio de Sierra Nevada (OSN)\footnote{The Observatorio de Sierra
  Nevada is operated by the Consejo 
Superior de Invetigaciones Cient\'{\i}ficas through the Instituto de Astrof\'{\i}sica de 
Andaluc\'{\i}a (Granada, Spain).} (Granada, Spain) on 2008 June 24. A RoperScientific VersArray 
with 2048$\times$2048 pixels of 0.232$\times$0.232\,arcsec$^2$ each was used as detector. Images were obtained 
through three narrow-band filters in the light of H$\alpha$ (FWHM =
10\,{\AA}), [N\,{\sc ii}]$\lambda$6583 
(FWHM = 10\,{\AA}), and [O\,{\sc iii}]$\lambda$5007 (FWHM = 50\,{\AA}). Exposure time was 1800\,s 
for each filter. Seeing was $\simeq$ 1.7\,arcsec during the observations. The images were reduced 
following standard procedures within the {\sc midas} package.

\begin{table}
 \centering  
\caption{Dereddened line intensities in IRAS\,18197--1118 in units of 
$I_{\rm H\beta}$ = 100.0}                           
\begin{tabular}{lrc}
\hline
Line & $f$($\lambda$) & $I$($\lambda$)  \\     
\hline                        
He\,{\sc i}\,$\lambda$3889  & 0.223 & 58   $\pm$ 7 \\     

H$\gamma$\,$\lambda$4340    & 0.129 & 48.5 $\pm$ 3.9 \\  

H$\beta$\,$\lambda$4861     & 0.000 & 100.0 $\pm$ 1.5 \\  

[O\,{\sc iii}]\,$\lambda$4959 & $-$0.023 & 413.0 $\pm$ 5.6 \\    

[O\,{\sc iii}]\,$\lambda$5007 & $-$0.034 & 1231.7 $\pm$ 13.1 \\   

[N\,{\sc ii}]\,$\lambda$5755 & $-$0.191 & 4.0 $\pm$ 0.2 \\ 

He\,{\sc i}\,$\lambda$5876  & $-$0.216 & 22.1 $\pm$ 0.3 \\ 

[O\,{\sc i}]\,$\lambda$6300 & $-$0.285 & 5.6 $\pm$ 0.1 \\   

[S\,{\sc iii}]\,$\lambda$6312 & $-$0.287 & 3.3 $\pm$ 0.1 \\ 

[O\,{\sc i}]\,$\lambda$6363 & $-$0.294 & 2.6 $\pm$ 0.1 \\ 

[N\,{\sc ii}]\,$\lambda$6548 & $-$0.321 & 28.9 $\pm$ 3.1 \\

H$\alpha$\,$\lambda$6563     & $-$0.323 & 315.2 $\pm$ 6.7 \\ 

[N\,{\sc ii}]\,$\lambda$6584 & $-$0.326 & 101 $\pm$ 3 \\

He\,{\sc i}\,$\lambda$6678  & $-$0.338 & 5.8  $\pm$ 0.1 \\   

[S\,{\sc ii}]\,$\lambda$6716 & $-$0.343 & 1.8 $\pm$ 0.1 \\

[S\,{\sc ii}]\,$\lambda$6731 & $-$0.345 & 4.3 $\pm$ 0.1 \\

He\,{\sc i}\,$\lambda$7065  & $-$0.383 & 13.7  $\pm$ 0.2 \\   

[Ar\,{\sc iii}]\,$\lambda$7135 & $-$0.391 & 40.7 $\pm$ 0.5 \\

He\,{\sc i}\,$\lambda$7281  & $-$0.406 & 1.2  $\pm$ 0.1 \\   

[O\,{\sc ii}]\,$\lambda$7325 & $-$0.411 & 16.3 $\pm$ 0.2 \\

[Ar\,{\sc iii}]\,$\lambda$7751 & $-$0.451 & 9.7 $\pm$ 0.1 \\

P\,20\,$\lambda$8392     & $-$0.509 & 0.16 $\pm$ 0.02 \\ 

P\,19\,$\lambda$8413     & $-$0.512 & 0.21 $\pm$ 0.02 \\ 

P\,18\,$\lambda$8438     & $-$0.516 & 0.91 $\pm$ 0.02 \\ 

P\,17\,$\lambda$8467     & $-$0.521 & 0.39 $\pm$ 0.02 \\ 

P\,16\,$\lambda$8502     & $-$0.526 & 0.72 $\pm$ 0.02 \\ 

P\,15\,$\lambda$8545     & $-$0.532 & 0.84 $\pm$ 0.02 \\ 

P\,14\,$\lambda$8598     & $-$0.540 & 0.98 $\pm$ 0.02 \\ 

P\,13\,$\lambda$8665     & $-$0.550 & 1.17 $\pm$ 0.02 \\ 

P\,12\,$\lambda$8750     & $-$0.562 & 1.46 $\pm$ 0.02 \\ 

P\,11\,$\lambda$8863     & $-$0.578 & 1.79 $\pm$ 0.03 \\ 

P\,10\,$\lambda$9015     & $-$0.599 & 1.06 $\pm$ 0.02 \\ 

[S\,{\sc iii}]\,$\lambda$9069 & $-$0.606 & 29.1 $\pm$ 0.3 \\ 

P\,9\,$\lambda$9229      & $-$0.612 & 2.79 $\pm$ 0.04 \\ 

\hline
log\,$F_{\rm H\beta}$ & $-$13.95 & (erg\,cm$^{-2}$\,s$^{-1}$)  \\
$c_{\rm H\beta}$     & 3.37 &    \\
\hline
\end{tabular}

\end{table}

\section{Results}

\begin{figure*}
   \begin{center}
   \includegraphics[angle=0,width=180mm,clip=]{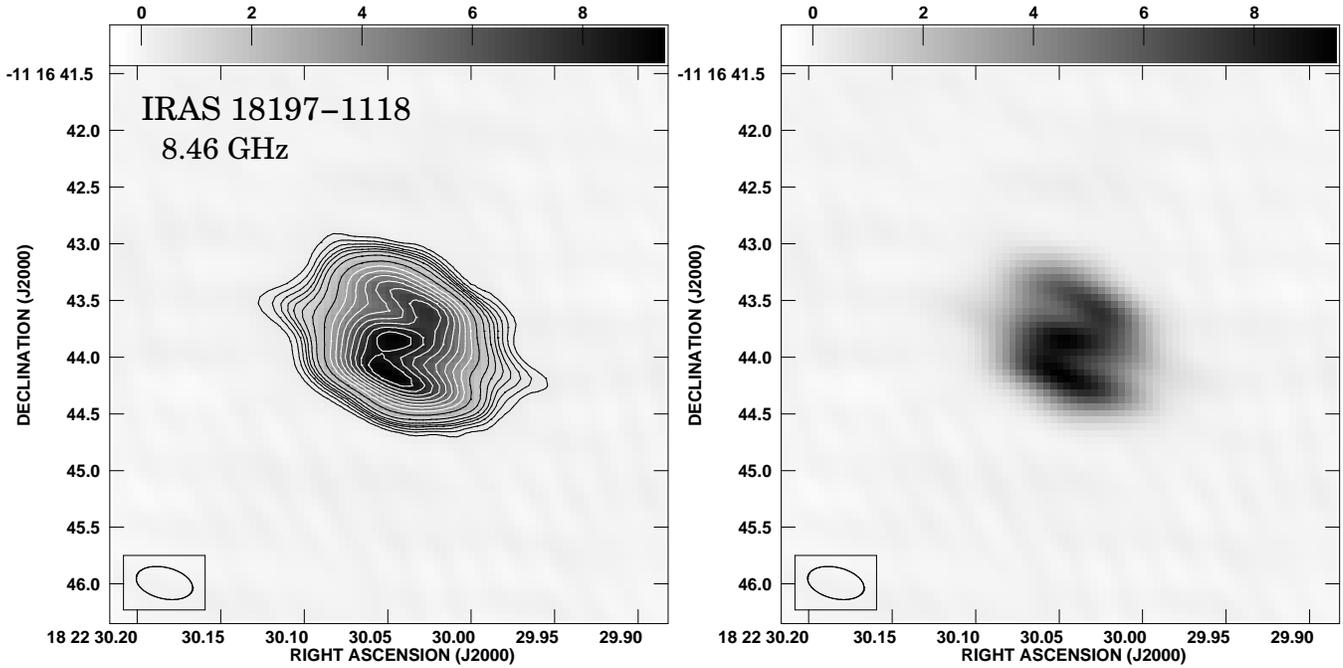}
      \caption{VLA image of IRAS~18197$-$1118 at 8.46 GHz in contours and grey
        scale (left panel), and grey scale (right panel). The contours are -4,
        4, 6, 8, 10, 12,  15, 20, 30, 40, 50, 60, 70, 80, 90, 100,  and 110
        times 0.079 mJy beam$^{-1}$, the rms noise of the image. The greyscale
        values are given in the upper wedge in units of mJy\,beam$^{-1}$. The
        synthesized beam, with dimensions of 0.51$\times$0.28\,arcsec$^2$  and
        major axis at position angle $77^\circ$, is shown in the bottom
        left corner of each panel.}
    \end{center}
   \end{figure*}

\subsection{The optical spectrum}

Figure\,1 presents the blue and red CAFOS spectra of
IRAS\,18197$-$1118. Hydrogen and neutral helium recombination 
emission lines and several forbidden emission lines in different excitation states can be 
recognized, as well as a faint nebular continuum. We note that the He\,{\sc
  ii}\,$\lambda$4686 and [O\,{\sc iii}]$\lambda$4363 emission lines are not
detected in our spectrum. In particular, an upper limit of
  3.4$\times$10$^{-16}$\,erg\,cm$^{-2}$\,s$^{-1}$ is obtained for the observed
flux in the [O\,{\sc iii}]$\lambda$4363 emission line. The observed 
H$\beta$ flux is $\simeq$ 1.12$\times$10$^{-14}$\,erg\,cm$^{-2}$\,s$^{-1}$. 

We use the H$\alpha$/H$\beta$ and H$\beta$/H$\gamma$ observed flux ratios to
calculate the logarithmic extinction coefficient 
($c_{\rm H\beta}$) assuming case B recombination, theoretical H$\alpha$/H$\beta$
and H$\beta$/H$\gamma$ line
ratios of 2.85 and 2.13, respectively (Brocklehurst 1971), and the extinction curve by Seaton (1979). 
We obtain $c_{\rm H\beta}$ $\simeq$ 3.50 from the H$\alpha$/H$\beta$ ratio 
and $\simeq$ 3.25 from the H$\beta$/H$\gamma$ ratio. Besides, the flux
density at 8.46\,GHz and the H$\beta$ flux can be used to derive $c_{\rm H\beta}$ $\simeq$ 3.55 
(see Pottasch 1984), similar to the values obtained from the Balmer
decrement. In the following, we will adopt $c_{\rm H\beta}$ = 3.37 as the mean
value from the Balmer lines and note that the main conclusions of this work 
do not depend on the assumed value of $c_{\rm H\beta}$ in the 3.25$-$3.55 range.

Table\,1 presents the dereddened emission line intensities obtained from the
spectrum, as well as their Poissonian errors. 
The [N\,{\sc ii}]$\lambda\lambda$6548,6583/H$\alpha$ and [S\,{\sc ii}]$\lambda\lambda$6716,6731/H$\alpha$ 
line intensity ratios of $\simeq$ 0.41 and $\simeq$ 0.02,
respectively, place IRAS\,18197--1118 in the PN region of the [N\,{\sc
  ii}]/H$\alpha$ vs. [S\,{\sc ii}]/H$\alpha$ diagram 
(see, e.g., Frew \& Parker 2010), confirming its PN nature. These two line
intensity ratios, the [O\,{\sc iii}]$\lambda\lambda$4959,5007/H$\beta$ 
line intensity ratio of $\simeq$ 16 (Table\,1), and the absence of He\,{\sc
  ii}$\lambda$4686 line emission indicate a medium-excitation PN.

\begin{figure}
   \begin{center}
   \includegraphics[width=70mm,clip]{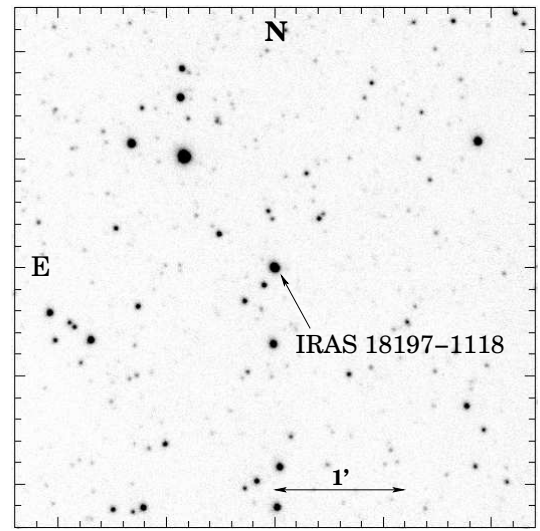}
   \caption{Identification chart of IRAS\,18197--1118 (arrowed) derived from the OSN [N\,{\sc ii}] image. }
    \end{center}
    \end{figure}

\subsection{Morphology}

Figure\,2 shows reproductions of the image of IRAS\,18197$-$1118 at 8.46 GHz
in which the nebula is resolved. The brightest nebular regions present a 
Z-shaped morphology (Fig.\,2, right) formed by two arcs and a bright, compact 
central region. When lower intensity levels are considered (Fig.\,2, left),
the arcs trace the brightest parts of an elliptical shell with a size of $\simeq$
2.7$\times$1.6\,arcsec$^2$ and the major axis oriented at PA $\simeq$
73$^{\circ}$. The twisted appearance of the arcs delineates a point-symmetry in 
the elliptical shell, that is similar to that observed in other point-symmetric elliptical PNe (e.g.,  
Miranda et al. 1997; 
Guerrero et al. 2001). Two faint, elongated bipolar protrusions, separated
by $\simeq$ 2.2\,arcsec, are observed along the major nebular axis of the elliptical shell. Their morphology 
suggests that they could be jet-like features. However, as the protrusions are elongated in 
a similar direction to that of the beam (Fig.\,2), the reality of
these features needs confirmation. The bright central region seems to be elongated
close to the east-west direction, although it is not resolved by the
observations (size $\la$ 0.5\,arcsec) and analyzing its morphology requires higher spatial
resolution. We note that the detection of the central region at radio
  continuum wavelengths implies a nebular structure and not emission from the
  central star. The existence of this region clearly 
shows that IRAS\,18197$-$1118 does not exhibit a hollow shell as expected in PNe, indicating that a complex mass 
loss history, possibly related to several ejection events, has been implied in the formation of this object. 

Figure\,3 shows an identification chart for IRAS\,18197--1118 based on the OSN
[N\,{\sc ii}] image. The object appears relatively bright in the three filters
(H$\alpha$ and [O\,{\sc iii}] images not shown here), although no internal
structure can be recognized at the (relatively low) spatial resolution of our
images, but only a stellar-like object. 

\subsection{Kinematics}

The high-resolution, long-slit spectra do not resolve details of the 
spatial structure of IRAS\,18197--1118, given the inadequate spatial 
resolution for such a small object. Thus, the spectra do not allow us to clarify whether 
the bipolar protrusions (Fig.\,2) are real, because faint emission from these
features could be superposed by the stronger one from the elliptical shell.

\begin{figure}
   \begin{center}
   \includegraphics[angle=-90,width=70mm,clip]{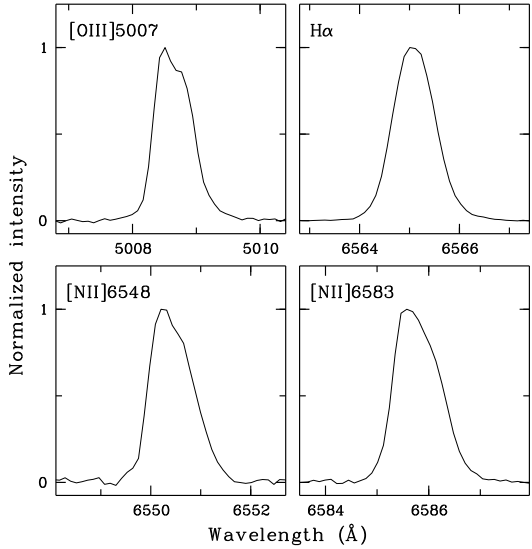}
   \caption{Normalized emission line profiles derived from the high-resolution, long-slit spectra.}
    \end{center}
    \end{figure}

Figure\,4 shows the integrated spectral profiles of the four observed emission 
lines. The H$\alpha$ emission line presents a single peaked, symmetric
profile, while the [N\,{\sc ii}] and [O\,{\sc iii}] emission line profiles are
asymmetric, suggesting that they are composed by at least two partially resolved velocity 
components. By using a two component Gaussian line fit, we 
obtain a radial velocity separation between the two velocity components of
$\simeq$ 26\,km\,s$^{-1}$ in [N\,{\sc ii}] and $\simeq$ 22\,km\,s$^{-1}$ in
[O\,{\sc iii}], which would imply a nebular expansion velocity of $\simeq$ 13
and $\simeq$ 11\,km\,s$^{-1}$ in [N\,{\sc ii}] and [O\,{\sc iii}],
respectively. However, because the emission lines are not well resolved, these values 
probably correspond to lower limits to the expansion velocity. We can consider the FWHM of 
the emission lines as more representative of the expansion velocity. In this case, and correcting 
of spectral resolution (\S\,2.2), we obtain an expansion velocity of $\simeq$ 22 and $\simeq$ 19\,km\,s$^{-1}$ in [N\,{\sc ii}] 
and [O\,{\sc iii}], respectively. These values should also be seen with caution because 
the FWHM does not discriminate velocity components that have not been spatially resolved and/or move almost 
perpendicular to the line of sight. In any case, with an expansion velocity of 20\,km\,s$^{-1}$ and the deconvolved 
size (0.85\,arcsec, see above), a crude estimate for the kinematical age of $\simeq$ 100$\times$$D$[kpc]\,yr is obtained, 
which, at a distance $D$ of 6\,kpc (see \S\,3.4), results to be $\simeq$ 600\,yr, pointing out to a relatively young PN. 
Finally, from the centroid of the emission line profiles we obtain a systemic velocity $V_{\rm LSR}$ = +95$\pm$1.5\,km\,s$^{-1}$ 
(V$_{\odot}$ = +80$\pm$1.5\,km\,s$^{-1}$) for IRAS\,18197--1118. 

\subsection{Physical conditions and chemical abundances}

The [S\,{\sc ii}]$\lambda$6716/$\lambda$6730 line intensity ratio of $\simeq$
0.42 (Table\,1) is at the lower sensitivity limit for electron density ($N_{\rm e}$)
measurement, indicating $N_{\rm e}$ $>$ 2$\times$10$^4$\,cm$^{-3}$. The
electron density and ionized mass ($M_{\rm i}$)  can be obtained from the radio continuum
emission following the formulation by Mezger \& Henderson 
(1967) for optically thin emission (see also G\'omez, Rodr\'{\i}guez \& Loinard 2013).
With the flux at 8.46\,GHz, the deconvolved radius at 8.46\,GHz, and an electron temperature 
$T_{\rm e}$ = 10000\,K, we obtain $N_{\rm e}$ $\simeq$
  8.23$\times$10$^4$$\times$$D$[kpc]$^{-0.5}$\,cm$^{-3}$ and $M_{\rm i}$
  $\simeq$ 2.4$\times$10$^{-2}$$\times$$D$[kpc]$^{2.5}$\,M$_{\sun}$. The distance to the nebula 
is involved in the calculations of these two parameters but is unknown for
IRAS\,18197$-$1118. We use the statistical distance scale by Zhang (1995) to
obtain a value of $\simeq$ 6\,kpc that will be used through the paper (see
also \S\,4). With this distance, the electron density and ionized mass are 
3.36$\pm$0.50$\times$10$^4$\,cm$^{-3}$ and
2.1$\pm$0.3$\times$10$^{-2}$\,M$_{\sun}$, respectively. The electron density 
is relatively high and compatible with the lower limit indicated by the
[S\,{\sc ii}] emission lines. The ionized 
mass is relatively small. The values of these two parameters suggest that
IRAS\,18917$-$1118 is a young PN.

\begin{table}
\caption{Ionic abundances relative to H$^+$ in IRAS\,18197$-$1118.}      
\centering    
\begin{tabular}{lc}
\hline         
Ion$^a$  &  Ionic abundance      \\
\hline                        
He$^+$         & 0.145$\pm$0.002                  \\  
O$^0$          & 8.5$\pm$1.7$\times$10$^{-6}$       \\
O$^+$          & 4.1$\pm$1.2$\times$10$^{-5}$       \\
O$^{2+}$       & 3.1$\pm$0.5$\times$10$^{-4}$       \\
N$^+$          & 2.0$\pm$0.4$\times$10$^{-5}$        \\
S$^+$          & 7.2$\pm$1.1$\times$10$^{-6}$         \\
S$^{2+}$       & 5.8$\pm$0.7$\times$10$^{-6}$       \\
Ar$^{2+}$      & 2.9$\pm$0.3$\times$10$^{-6}$    \\
\hline
\end{tabular} 

$^a$ For ions with more than one transition, an intensity-weighted average has been used.

\end{table}

\begin{table}
\centering  
\caption{Elemental abundances in IRAS\,18197$-$1118.}
\begin{tabular}{lccc}
\hline
Element ratio   & Abundance \\
\hline
He/H                     &  0.145$\pm$0.002    \\
O/H                      &  3.5$\pm$0.5$\times$10$^{-4}$         \\
N/H                      &  1.7$\pm$0.6$\times$10$^{-4}$       \\
S/H                      &  9.7$\pm$3.7$\times$10$^{-6}$        \\
Ar/H                     &  5.4$\pm$0.3$\times$10$^{-6}$        \\
N/O                      &  0.48$\pm$0.17     \\

\hline
\end{tabular}
\end{table}

The detection of the auroral and nebular [N\,{\sc ii}] emission lines
(Table\,1) allows us to obtain the electron temperature $T_{\rm e}$([N\,{\sc
  ii}]). We have used the task temden in {\sc iraf} and the value obtained for
$N_{\rm e}$ to derive $T_{\rm e}$([N\,{\sc ii}]) =
11160$\pm$575\,K. We note that, because of the weak dependece of $N_{\rm e}$ and $M_{\rm i}$ on $T_{\rm e}$ 
($N_{\rm e}$ and $M_{\rm i}$ $\propto$ [$T_{\rm e}$/10$^4$]$^{0.175}$, see, e.g., G\'omez et al. 2013), the use of  $T_{\rm e}$ = 
10000\,K or 11160\,K to calculate $N_{\rm e}$ and $M_{\rm i}$ is not critical.

Ionic and elemental abundances have been obtained for the values of $N_{\rm
  e}$ and $T_{\rm e}$([N\,{\sc ii}]) quoted above. For the ionic 
abundances we used the task ionic of {\sc iraf} and they are listed in
Table\,2. Total elemental abundances have been 
calculated using the icf method by Kingsburgh \& Barlow (1994). For the helium
abundance, we follow the formulation by Clegg (1987). The elemental abundances
are listed in Table\,3. In Figure\,5 we show a 12+log(S/H) vs. 12+log(O/H)
abundance diagram to analyze the Peimbert type (Peimbert 1990) of IRAS\,18197$-$1118.

The position of IRAS\,18197$-$1118 in Fig.\,5 rules out a type\,IV (halo) PN 
while type\,I and type\,II PNe cannot be distinguished in this diagram. The
high helium abundance (He/H $\simeq $0.14) indicates a type\,I PNe. The
N/O abundance ratio of $\sim$ 0.48 (Table\,3) seems compatible with a type\,I
PN (N/O $\geq$ 0.5, Peimbert 1978), although the involved errors do not allow us to draw a definitive
conclusion. We note that a N/O abundance ratio of 0.65--0.8 has been considered by 
Kingsburgh \& Barlow (1994) and Henry, Kwitter \& Balick (2004) as a lower limit for 
type\,I PNe. With this criterion, the N/O abundance ratio classifies
IRAS\,18197$-$1118 as a type\,II PN. This classification 
is compatible with the nitrogen abundance (Table\,3) that is much lower than that observed in 
type\,I PNe. Chemical abundances in PNe are related to the evolution of the
progenitor star and, in particular, to its mass. In this respect, the double
classification of IRAS\,18197$-$1118 presents some contradictions. While a
type\,II classification suggests a relatively low-mass progenitor, a type\,I
one requires a relatively massive progenitor (e.g., Corradi \& Schwarz 1995).
It is interesting to mention that other type\,II PNe also show a high helium
abundance typical of type\,I PNe  (see Henry et al. 2004).

\begin{figure}
   \begin{center}
   \includegraphics[width=70mm,clip]{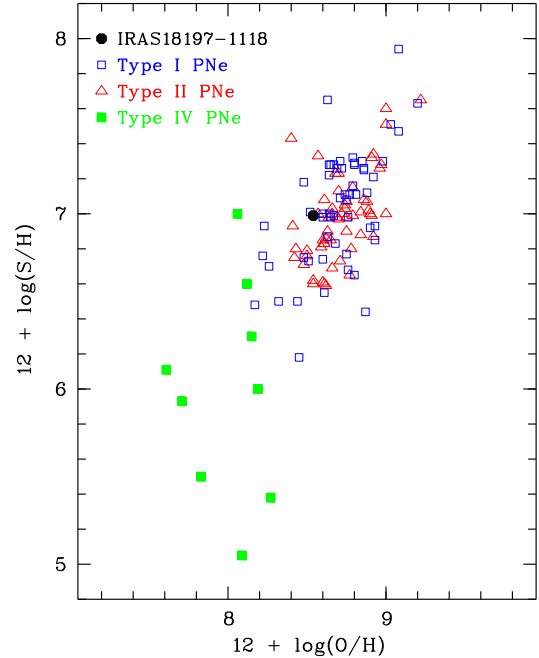}
   \caption{Plot of 12+log(S/H) vs. log(OH) for the four Peimbert types of
     PNe. The data are taken from Maciel \& K\"oppen (1994), 
 Howard, Henry \& McCartney (1997), and Pereira \& Miranda (2007). The position of 
IRAS\,18197--1118 is indicated (see Table\,3) (see the electronic version for a colour version of 
this figure). }
    \end{center}
    \end{figure}

To obtain more precise information about the properties of the
  IRAS\,18197$-$1118 progenitor, we have compared the obtained abundances
  (Table\,3) with those predicted from the evolutionary models by Marigo
  (2001) and Karakas (2010). According to Marigo's (2001) models, the obtained N and O
  abundances suggest a 1.3--1.5\,M$_{\odot}$ progenitor, although with a
  different metallicity, Z = 0.019 for N, and Z = 0.008 for O. However, the
  observed He abundance requires a progenitor mass $>$ 4\,M$_{\odot}$ for any
  metallicity. Using the Karakas' (2010) models, the obtained N abundance
  points out to a 2.5\,M$_{\odot}$ progenitor with Z = 0.02, while the
  obtained O abundance is not well reproduced, and only models
  with 2\,M$_{\odot}$ produce approximated values. As for the He
  abundance, Karakas' models predict (much) lower values than the obtained
  one, even considering a 6\,M$_{\odot}$ progenitor. We note the existence of
  noticeable discrepancies between the predictions of the two models, which could be
  attributable to the many different assumptions included in each model. However, both
  models coincide in that the observed He abundance requires a much more
  massive progenitor than indicated by the observed N and
  O abundances. In the case of IRAS\,18197$-$1118 this could be related to the
  presence of two different nebular regions, the elliptical shell and the
  central region, each with different abundances (see \S\,4).

\section{Discussion}

Our data have confirmed that IRAS18197--1118 is a true PN and have allowed
us to deduce many of its properties. The values obtained for the
electron density, ionized mass, and kinematical age
indicate a relatively young PN. Particularly interesting is
the high extinction towards the object. The value of $c_{\rm H\beta}$ $\simeq$
3.25$-$3.55 found in IRAS18197--1118 is high for PNe and higher than the
$c_{\rm H\beta}$ values found in the large sample of PNe analyzed by Stasi\'nska et
al. (1992).

IRAS18197--1118 is also atypical in that it contains a compact central 
  nebular region inside an extended elliptical shell. There are only a few PNe 
that show a compact central nebula besides an extended (and typical) elliptical or bipolar shell, 
as, e.g., M\,2-29, EGB\,6, M\,2-48 (Gesiki et al. 2010 and references therein), IC\,4997 (Miranda \& Torrelles 1998), 
and KjPn\,8 (V\'azquez, Kingsburgh \& L\'opez 1998). The existence of the central nebular region and the 
elliptical shell indicates that
  the formation of IRAS18197--1118 has been complex, involving at least two different
  mass ejection events. Moreover, it is reasonable to assume
  that the two ejections are not coeval but the formation of the elliptical shell has
  preceded that of the central region. If so, the chemical abundances (and
  physical conditions) in the central region could be different from those in
  the elliptical shell. Because the object is not spatially resolved in the
  spectra, the contribution of each structure to the (integrated) emission
  line intensities cannot be determined, and the deduced abundances (and
  physical conditions) should
  represent an (unknown weighting) average over the nebula. It is interesting
  to compare IRAS18197--1118 with Abell\,30 and
  Abell\,58, although most probably the formation of IRAS18197--1118 has
  nothing to do with the born-again phenomenon of Abell\,30 and
  Abell\,58. These two PNe present old
  and much younger nebular ejections that contain completely different chemical
  abundances from each other (Guerrero \& Manchado 1996). One might wonder
  about what chemical abundances would be obtained in Abell\,30 and Abell\,58
  if they were spatially unresolved, and whether a comparison of these
  abundances with evolutionary models would indeed provide coherent and
  realistic information about their progenitor. It should be emphasized that
  evolutionary models, as those mentioned above, do not incorporate multiple,
  episodic ejections in the formation of PNe, as it is observed in many of
  these objects. Therefore, a proper comparison of observed and model
  abundances seems to require previous information about the morphology of the
  object. This may be particularly relevant for compact PNe and crucial for
  those presenting ``peculiarities'' in their chemical abundances, as IRAS18197--1118.

The Galactic coordinates and the estimated (statistical) distance of 6\,kpc
place IRAS18197--1118 in the inner Galactic disc. In fact, the radial velocity of 
the object ($V_{\rm LSR}$ $\simeq$ +95\,km\,s$^{-1}$) fits very well in the distribution of 
radial velocity vs. Galactic longitude observed in inner-disc PNe (see Chiappini et al. 
2009). Moreover, the radio continuum flux at 5\,GHz of
IRAS\,18197--1118 ($\simeq$ 61\,mJy, Urquhart et al. 2009) is lower 
than 100\,mJy, and the size of the object ($<$ 2.7 arcsec, see above) is much
smaller than 10\,arcsec, two criteria to be fulfilled  by PNe located in the inner Galaxy
(Stasi\'nska et al. 1991; Chiappini et al. 2009). We also note that the high extinction towards
the object points to a relatively large distance. Values of $c_{\rm H\beta}$
comparable to or (much) higher than that observed in IRAS\,18197--1118 are
found in a number of inner-disc and bulge PNe, at distances $>$ 5\,kpc (e.g., Van de
Steene \& Jacoby 2001; Exter, Barlow \& Walton 2004), whereas PNe 
at distances $\la$ 4\,kpc usually present $c_{\rm H\beta}$ $\la$ 2 (see,
e.g., Cappellaro et al. 2001; Giammanco et al. 2010). As for the abundances
and abundance ratios, the values found in IRAS\,18197--1118 are within
the ranges observed in PNe in the inner Galactic regions (e.g., Henry
et al. 2004; Exter et al. 2004; Cavichia et al. 2010). 

\begin{table}
\centering  
\caption{Electron density, electron temperature, ionized mass, and elemental 
abundances in IRAS\,18197$-$1118 
for distances of 2 and 10\,kpc.}
\begin{tabular}{lcc}
\hline

Parameter   &  $D$ = 2\,kpc   &  $D$ = 10\,kpc   \\

\hline

$N_{\rm e}$ (cm$^{-3}$)                & 5.82$\pm$0.87$\times$10$^{4}$    & 2.60$\pm$0.39 $\times$10$^{4}$ \\

$T_{\rm e}$([N\,{\sc ii}]) (K)         &  9350$\pm$415                   & 12000$\pm$700 \\

$M_{\rm i}$ (M$_{\odot}$)               & 1.3$\pm$0.2$\times$10$^{-3}$   & 7.5$\pm$0.9$\times$10$^{-2}$    \\

\hline

He/H                    & 0.144$\pm$0.003    & 0.146$\pm$0.002 \\

O/H                     & 7.0$\pm$1.0$\times$10$^{-4}$        & 2.8$\pm$0.5$\times$10$^{-4}$        \\

N/H                     & 2.5$\pm$0.9$\times$10$^{-4}$         & 1.3$\pm$0.5$\times$10$^{-4}$       \\

S/H                     & 1.4$\pm$0.5$\times$10$^{-5}$       &  8.1$\pm$3.8$\times$10$^{-6}$       \\

Ar/H                    & 8.2$\pm$0.4$\times$10$^{-6}$          & 4.7$\pm$0.3$\times$10$^{-6}$       \\

N/O                     & 0.35$\pm$0.12      & 0.46$\pm$0.19  \\

\hline
\end{tabular}
\end{table}

A critical parameter in the analysis carried out above is the distance to
IRAS\,18197--1118. Through this paper we have considered 6\,kpc, which has been 
estimated from the statistical distance scale by Zhang (1995). Similar values for 
the distance of $\simeq$ 7\,kpc and $\simeq$ 6.3\,kpc are found with the distance scales 
by Van de Steene \& Ziljstra (1995) and Bensby \& Lundstr\"om (2001),
respectively, providing support for the distance used in this
paper. Nevertheless, some results suggest that statistical distances may not be
suitable to analyze individual PNe (Tafoya et al. 2011; Miranda et
al. 2012). Therefore, we consider it interesting to check whether the main
conclusions of this investigation might critically depend on the distance value. To do this, we
have considered a small and a large distance, namely, 2 and 10\,kpc (i.e.,
6$\pm$4\,kpc), to obtain the physical conditions and elemental abundances in IRAS\,18197--1118, 
which should be compared with those obtained for 6\,kpc. 
Table\,4 lists the electron density, electron temperature, ionized mass, and 
elemental abundances for 2 and 10\,kpc, derived using the same procedures as done  
for 6\,kpc. In the 2--10\,kpc distance range, the electron density and ionized
mass are relatively high and low, respectively, and compatible with 
a relatively young PN. The helium abundance does not show a strong dependence on the distance 
and indicates a type\,I PN. The N/O abundance ratio is comparable to or smaller than that
obtained for 6\,kpc, and, in any case, lower than the
value of 0.65--0.8, suggesting a type\,II PN (see above). Similarly, the nitrogen
abundance remains much lower than typically observed in type\,I PNe, also 
suggesting a type\,II classification. A comparison of observed and model abundances, as 
done for 6\,kpc (\S\,3.4), also shows that the observed He abundance requires
a more massive progenitor than indicated by the observed N and O abundances. 

At a distance of about 10\,kpc, IRAS\,18197--1118 would still be located in
the inner-disc. For distances of, say, $\la$ 4\,kpc, IRAS\,18197--1118 would
be outside of the inner Galactic regions. However, the criteria of the radio
continuum flux at 5\,GHz and angular size of the object (see above) exclude
$\sim$ 90--95\% of PNe that are observed towards the inner Galaxy but do not
belong to it (see Stasi\'nska et al. 1991). Therefore, the probability that IRAS\,18197--1118 
is located at $\la$ 4\,kpc is small. Moreover, as already mentioned, PNe
  at $\la$ 4\,kpc, present values of $c_{\rm H\beta}$ smaller than that found in
  IRAS\,18197--1118. These comments argue in favor of 
IRAS\,18197--1118 being located in the inner-disc and suggest that 6\,kpc may be an
approximate lower limit to the distance to the object.

\section{Conclusions}

We have presented an analysis of optical intermediate- and high-resolution
spectra, 8.64\,GHz radio continuum data, and narrow-band optical images of
IRAS\,18197--1118, a PN candidate whose nature, spectral properties 
and morphology had not been investigated before in detail. Our conclusions can
be summarized as follows: 

\begin{itemize}

\item The spectra confirm the PN nature of IRAS\,18197--1118. A particularly high extinction 
($c_{\rm H\beta}$ $\simeq$ 3.25--3.55) is obtained towards the object. 

\item The image at 8.64\, GHz reveals a small (size $\simeq$ 2.7$\times$1.6
  arcsec$^2$), point-symmetric elliptical PN, a bright compact central
  region, and possible bipolar jet-like features, indicating several ejection
  events. The presence of a compact central nebula is unusual among PNe. 

\item An expansion velocity of $\simeq$ 20\,km\,s$^{-1}$, a systemic velocity (LSR) 
of $\simeq$ +95\,km\,s$^{-1}$, and a kinematical age of $\simeq$ 100$\times$$D$[kpc]\,yr are 
obtained for the object. The kinematical age suggests a relatively young PN. 

\item A relatively high electron density ($\simeq$
    3.4$\times$10$^4$\,cm$^{-3}$) and a relatively small ionized mass ($\simeq$
    2.1$\times$10$^{-2}$\,M$_{\sun}$) are derived from the
  8.64\,GHz radio continuum data for a distance of 6\,kpc estimated from a statistical
  distance scale. Electron density and ionized mass also suggest a relatively young PN. 

\item A high He abundance is obtained but N is not enriched. The observed
  abundances cannot be reproduced in a consistent manner with standard evolutionary
  models, suggesting that the central region and the elliptical shell present
  different abundances. 
.
\item The Galactic coordinates, radio continuum flux at 5\,GHz, small 
  angular size, systemic velocity and estimated distance of 6\,kpc are
  consistent with IRAS\,18197--1118 being an inner-disc PNe. The high extinction 
towards the object is also compatible with a relatively large distance.

\item The main conclusions of this work do not critically depend on the distance to
  IRAS\,18197--1118. Nevertheless, the properties of the object argue in favor
  of a distance $\ga$ 6\,kpc. 

\end{itemize}

\section*{Acknowledgments}

We are very grateful to our referee, A. Zijlstra,  for valuable comments that have 
improved the presentation and interpretation of the data. We thank Calar Alto Observatory for 
allocation of director's discretionary time to this programme. We are very grateful to the staff on 
Calar Alto for carrying out the observations. Discussions with A. Garc\'{\i}a-Hern\'andez and 
M.A. Guerrero are warmly acknowledged. We thank the staff of OAN-SPM, in particular to 
Mr. Gustavo Melgoza-Kennedy, for assistance during observations. LFM
acknowledges partial support from Spanish MICINN AYA2011-30228-C3-01 grant (co-funded by FEDER funds). 
RV acknowledges support from grant UNAM-DGAPA-PAPIIT IN107914. 

The investigation of IRAS\,18197--1118 was suggested by Yolanda G\'omez some
years ago, when she found and analyzed the object in the VLA archive. We would like to
dedicate this paper to the memory of Yolanda who passed away on 2012 February 16.

\end{document}